# An Acataleptic Universe

Philip Gibbs

John Wheeler advocated the principle that information is the foundation of physics and asked us to reformulate physics in terms of bits. The goal is to consider what we know already and work out a new mathematical theory in which space, time and matter are secondary. An application of the converse of Noether's second theorem to the holographic principle implies that physics must have an underlying hidden symmetry with degrees of symmetry that match physical degrees of freedom in order to account for the huge redundancy of information in the interior of a black-hole. I have been working on a theory that builds infinite dimensional symmetries using layers of quantisation from information as suggested by Wheeler's contemporary Carl von Weizsäcker. Necklace Lie algebras are the mathematical objects and iterated integration can be used to show how a continuum background can emerge from their structure. The logic suggests the conclusion that wheeler was right when he proclaimed "It from Bit"

## *Introduction*

John Wheeler's apothegm "It From Bit" was the title of a 1989 essay in which he elucidated his vision that the physical laws of the universe might arise from the dynamics of "yes" or "no" valued binary digits [1]. Nothing else in our multifarious universe is fundamental, he conjectured. **Space and time, locality and causality, all matter and all processes are emergent phenomena from an ethereal pregeometry of quantum information**.

Wheeler was not alone. 1989 was the same year that Carl Friedrich von Weizsäcker became a Templeton Prize laureate for his foundational work at the boundary of physics and philosophy. Weizsäcker is celebrated amongst other things for his pre-war work on the nuclear processes that light the stars. From the 1950s he embarked on a courageous program to reconstruct physics starting from a single ur-alternative through a process of multiple quantisation [2]. The first quantization gave a single structure based on the group SU(2) that can encode the spin of a quark or electron. In modern terms we would call it a qubit. A further quantisation leads to the symmetries of 4 dimensional space-time in which a single particle moves. With yet further quantisations the single qubits become many qubits and so he hoped the physics of quantum field theory could emerge.

**Weizsäcker's ur-theory of Multiple Quantisation was more a dream of hope than the foundations for a system of real physics**. It grew from the observation that first quantisation of a single particle's dynamics is followed by a second quantisation to derive the quantum field theory of multi-particle systems, but modern field theorists have criticised this approach calling first quantisation a mistake and second quantisation a misnomer. In today's physics the Dirac equation along with Maxwell's equations form the classical system of field equations that are quantised just once to form the final theory of quantum electrodynamics. This has been extended with non-abelian gauge theories and the Higgs mechanism to define the successful standard model of particle physics. That said, who cannot see that there is something inherently quantum about the Dirac equation already as a "classical" field? Planck's constant appears in the equation and the spin half wave function is already

modelled by its four component fields. Most of Bohr's atom can already be understood from this equation before the second quantisation is applied. Something must be right about Weizsäcker's theory, but he was just too far ahead of his time to find the right formulation.

**Multiple quantisation seems to embody the acataleptic philosophy of ancient Greeks such as Carneades and Arcesilaus**. As skeptics they opposed the assertions of absolute truth made by the stoics two centuries before Christ. In their school of thought everything was fundamentally uncertain, even the degree of uncertainty itself. The reality of such uncertainty is borne out today by the laws of quantum mechanics that replace every classical variable with a wave function from which only the probability of any outcome can be predicted. **With multiple quantisation the values of the probabilities themselves are replaced with further wave-functions ad infinitum. In such a world, can we hope to determine anything?**

My claim which I try to justify in this essay is that this is indeed the correct way to understand the universe. **From layers of quantum uncertainty built upon fundamental information there is hope that spacetime and matter emerge in a natural way**. The key is the mathematics of information redundancy which brings symmetry through algebraic geometry. In particular I will outline the rise of causality from the holographic principle and the emergence of smooth spacetime from necklace algebras through iterated integration.

## *Holography and the Power of Consistency*

So let us assume – as a working hypothesis at least – that quantum information is fundamental and all material entities including space-time are emergent. How can we hope to pursue this idea? Using philosophy alone to find the right dynamics is unlikely to succeed. Existence may not share our human philosophical prejudices for simplicity, symmetry or anything else. Some observational input on phenomenology of quantum gravity would help but for now everything we can measure is adequately explained by the physics of quantum mechanics, general relativity, thermodynamics, and the standard model of particle physics. The emergence of space and time are phenomena in a realm of physical extremes way beyond what these can tell us directly. New observations may come in time, but meanwhile we have to work from what we have.

And yet there is hope. In the past, theorists have jumped far beyond the present knowns to predict new experimental results by using simply the requirement for logical consistency when combining different areas of known physics. Maxwell predicted radio waves by combining the laws of electric and magnetic fields. Einstein predicted the bending of light round the Sun after searching for a theory of gravity that would be consistent with the principles of relativity that applied to electromagnetism. Dirac predicted antimatter from a combination of special relativity with quantum physics. Finally, the standard model was the solution of finding a quantum field theory with heavy gauge bosons and fermions constrained by the consistency requirement of renormalisability leading amongst other things to the successful prediction of the Higgs boson. Contrary to the popular portrayal, these predictions worked because they came about as requirements of consistency. Simplicity, symmetry and mathematical elegance played their part in finding the answer, but consistency was the real guiding principle that said they had to be right. **Bringing together different theories often forces almost unique conclusions. What can we discover now by combining general relativity, quantum mechanics and thermodynamics in the same way?**

One answer is the holographic principle which can be deduced from considerations of the thermodynamics of black holes. I will summarise the train of arguments briefly before extending further with some less well-known thoughts of my own.

Black holes are a clear prediction of general relativity. When enough matter is brought together in a volume of space then gravitation pulls them together until a region is cut-off from the rest of the universe by an event horizon from within which nothing can escape. There are no absolute proofs in science but observational evidence for the existence of black holes in our galaxy is highly convincing. Any objects in the observable universe can be expected to obey the laws of thermodynamics and black holes are no exceptions. The second law of thermodynamics tell us that entropy increases. Such a law cannot be fundamental because the all the known underlying laws have time reversal symmetry (or at least CPT symmetry) so anything that can run forwards can also run backwards. The second law is statistical in nature and is emergent, but it holds very well in everything we observe and this is enough to deduce conclusions that are fundamental.

When Jacob Bekenstein calculated the change of entropy as particles drop into a blackhole he was led to the inescapable conclusion that its entropy $S$ is given by the area of the event horizon $A$ in Planck units times a quarter of Boltzmann's constant $k$

$$S = \frac{k}{4} A$$

Stephen Hawking then applied quantum mechanics to black holes to show that they must have a temperature consistent with this entropy and must radiate at a specific temperature that decreases with increasing mass. **The result is a set of laws that a more complete theory of quantum gravity must explain and it is derived from generic arguments independent of any specific theory. That is the amazing power of consistency, and it is just the beginning**.

**Hawking knew that entropy is directly related to information**. This conclusion comes from the work of Claude Shannon who analysed the amount of information in a string of bits like the content of this essay when stored in a computer [3]. The text is 25000 characters long which can be stored in ASCII code using 200000 bits, but if that is run through a file compression tool such as gzip it will reduce to about 80000 bits which is a better indication of the real amount of information contained. A compression tool is any program that uses a clever algorithm to find patterns such as repeated sequences in the string of bits that can be exploited to encode it with fewer bits by removing the redundancy. Analysing all possible algorithms to find the best possible compression of a given text is an infinitely complex and insoluble problem, but Shannon realised that if you considered a statistical ensemble of possible strings where each one appeared with probability $p_i$ then the average number of bits of information $B$ required is given by the formula

$$B = -\sum_i p_i \log_2 p_i$$

This is similar to the formula Boltzmann used to derive the entropy for an ensemble of physical states which appear in a statistical physics system with probability $p_i$

$$S = -k \sum_i p_i \ln p_i$$

Following the work of Shannon, physicist Edwin Jaynes argued that this is more than just a similarity [4]. **It tells us that the number of bits of information $N$ without any redundancy in a physical system with entropy $S$ is given by**,

$$N = \frac{S}{k \ln 2}$$

This means that you could take the information in a black hole and spread it over the event horizon in such a way that each bit lives in an area of $A = \frac{4}{\ln 2} l_p{}^2$ where $l_p$ is the unit Planck length, a tiny distance of about $10^{-35}$ meters.

**Sometimes the most brilliant step towards a great discovery is asking the right question to begin with.** This was certainly the case with the black hole information paradox. Hawking realised that if you throw an object into a black hole then the information would be hidden from outside the event horizon. If the black hole is then left to evaporate into Hawking radiation **where would the information have gone?** The radiation is perfectly random and should not be able to reveal what was inside the black hole because information would have to travel faster than light to get out. If entropy is information then a procedure like this could reduce the amount of entropy in a closed system and contravene the second law of thermodynamics. In quantum terms a pure state evolves to a mixed state defying unitarity. **Hawking recognised that this is a fundamental question whose resolution demanded a consistent explanation that would tell us something deep about the foundations of physics**.

The next step in the argument waited two decades to emerge. Gerard 't Hooft reasoned that Hawking's information-loss paradox implied a holographic principle for the laws of physics. **The amount of information in any volume of space must be limited by the area of a surface that encompasses it**, otherwise you could throw in heavy matter to create a black hole around the volume and lose some of the information. Leonard Susskind then provided further arguments to back this up, showing that string theory could be consistent with such a holographic principle, Finally the idea became more widely accepted when Juan Maldacena showed that one version of string theory in 5 dimensional anti-desitter space fulfilled the holographic principle because the gravitational theory in the bulk is dual to a 4-dimensional conformal field theory of the boundary.

So by consistency arguments alone theorists had been able to argue that the laws of physics must be holographic in nature. According to Naïve expectations you would think that it would be possible to build information storage devices where the amount of data held is limited by the volume of the space they occupy, but in reality the information content is bounded by a much stricter limit given by the area of a surrounding surface. It is as if most of the information that should be stored in the quantum fields is in fact redundant so that it cannot contribute to the entropy. The argument for this is not watertight and is not backed up by any experiment so far, but it is based on consistency reasoning from the need to bring together the laws of gravitation, quantum theory and thermodynamics. Either it is correct or some other deeply held assumption must break down along the way. In my opinion the assumptions are good for the physical conditions in which they have been used. They may break down at a deeper level but the reasoning works and the holographic principle is something we must work with.

## Holographic Explanations and Complete Symmetry

How can the holographic principle actually work? Maldacena's AdS/CFT correspondence is not understood well enough to answer this question. Now I will add some new ideas of my own based on conservation laws to try to get an idea for how this can be answered.

The way conservation laws work in physics has been well understood since the work of Emmy Noether [5]. **There is a correspondence between symmetries and conservation laws embodied in two theorems and their converses as proven by Noether**. These were originally cast in the context of classical physics under the assumption that a principle of least action determined the dynamics of a system, but they have also been applied to quantum mechanical systems. The most basic example is energy conservation which is related to time symmetry. If a physical system does not have explicit time dependence then conservation of energy follows from Noether's first theorem.

Noether's work arose in the context of general relativity shortly after Einstein had formulated his gravitational field equations and derived an expression for energy conservation for gravity. Mathematician Felix Klein cast doubt on how this worked and told Einstein that his energy conservation law reduced to a trivial mathematical identity rather than a proper physical law. This was because his expression for energy and momentum currents in the gravitational field split into a sum of two parts. One part was zero everywhere due to the field equations and the other part was the divergence of an anti-symmetric tensor which must be conserved independently of the dynamics. David Hilbert sided with Klein and enlisted Noether to investigate. Noether derived her general theorems to back up the claim. Einstein did not have the mathematical sophistication to contradict their conclusions but he still felt he was right. It took several decades to resolve the question in favour of Einstein by showing that gravitational energy is carried in gravitational waves. **Energy conservation in general relativity is real, exact, non-trivial and important**.

The symmetry of general relativity is diffeomorphism invariance where each diffeomorphism is generated by a vector field $\xi^\mu$. In the case of energy conservation the field must be time-like so it generates a time translation. Noether's first theorem can be used to derive the corresponding conserved energy current. The traditional form of Noether's theorem requires that the action depends only on the field variables and their first derivatives, but in general relativity where the field variables are the metric tensor there is a dependency on the second derivatives. The usual procedure to work round this is to remove the second derivatives but this results in a non-covariant form for the energy current using pseudotensors. I prefer to generalise Noether's theorem to work directly with the second derivatives. This provides a local covariant expression for the current [6]

$$J^\nu(\xi^\mu) = \xi^\mu T_\mu{}^\nu - \frac{1}{\kappa}\left(\xi^\mu T_\mu{}^\nu G_\mu{}^\nu + \xi^\nu \Lambda\right) + K^{\mu\nu}{}_{;\mu}$$

$$K^{\mu\nu} = \frac{1}{2}(\xi^{\mu;\nu} - \xi^{\nu;\mu})$$

The price paid with this formulation is that the current has a direct dependency on field $\xi^\mu$ and its derivatives which tells us that energy is a relative quantity that cannot be separated from this field. When field equations are applied the first terms vanish leaving only the last. We find that the divergence of this current is zero because $K^{\mu\nu}$ is anti-symmetric.

$$J^\nu{}_{;\nu} = 0$$

As Klein said, the energy current is a sum of one term that is zero from the field equations and another whose divergence is identically zero, but this does not mean that energy conservation is trivial.

It means we can integrate over a volume of space to get a value for energy in terms of an expression integrated over the surface only. This is the energy analogue of Gauss's flux theorem for an electric charge which can be determined from the electric field on a surrounding surface. Some physicists like to say that this makes energy a non-local concept in general relativity but the better way to describe it is that **energy is holographic**. The energy within a volume of space can be determined by looking at the gravitational flux over the bounding surface. This is even true when the volume is the inside of a black hole. The energy contained in everything that has been thrown in can be determined from the outside. It is just the black hole mass. In addition we know its momentum and angular momentum. The same works with charges from any other gauge theory. **Information about the total electric charge, colour charge and weak-isospin charge thrown into a black hole is not lost**, but according to classical theory all other information is. This is the no-hair theorem for black holes.

The holographic principle tells us that there must be much more information available from a black hole than just these total charges. If they are lost at the level of classical physics they must be a quantum phenomenon. **But could it be that all the hidden information actually comes in the form of charges from gauge symmetries?**

In the twentieth century gauge symmetries were the angels of physics at the centre of all successful theories from general relativity to the standard model and beyond, but in the last decade physicists have been more disparaging, saying that **symmetry is just a kind of redundancy** that takes different forms in different dual versions of a theory. I disagree. I think that symmetry has only begun to reveal its true size and power. There is much more of it that lies hidden. **Redundancy is in fact the key ingredient of the holographic principle**. In the bulk of space physics is described by field variables packing the volume, but if the only real information can be contained on the surrounding surface then most of the bulk field variables must be redundant.

Noether's second theorem applies to the case of gauge symmetry and tells us that for every degree of gauge symmetry there must be an equation of redundancy in the field equations, but both the first and second theorem of Noether have converses. **If there is redundancy then there must also be gauge symmetry. This is an inescapable consequence of Noether's theorems**. Yang-Mills gauge theory only provides partial redundancy. For example, in electrodynamics we can gauge fix by setting the time component of the vector potential to zero to remove all gauge redundancy except a single global constant. The remaining three quarters of the vector potential field remains. For holography to work all field variables must be removed leaving just a number of holographic variables on the boundary surface. This requires that **the system of fields used to describe physics in the bulk must have what I call "complete symmetry"**. That is, one degree of symmetry for every field variable. Furthermore, if there are fermionic fields then this requires fermionic degrees of symmetry to make them redundant, i.e. **supersymmetry is required**.

In traditional supergravity the generators of the super-lie algebra are represented by spin half and vector fields, but the dynamic fields include spin zero, spin half, spin one, spin one and a half and spin two fields. The symmetry falls short of what is required for complete symmetry. However,

**higher spin gravity theories include an infinite tower of spins and it may be possible to realise complete symmetry**. String theory also has higher spin modes and although such huge forms of symmetry have not been recognised in superstrings I think that the same idea applies.

The lesson to be taken from holography is that there is a huge hidden symmetry in physics that nobody has yet appreciated. It may be only visible in an algebraic pregeometric theory from which space time emerges. **To understand the foundations of universal law we need to look at complex infinite dimensional symmetries and use the adjoint representation for fields so that every field variable corresponds to a degree of freedom making it redundant**. Nevertheless such fields can contain real information given by the quantised charges of the symmetry. This is the information from which physics emerges.

## *Necklace Lie Algebras and Iterated Integration*

In my work over the last twenty years I have explored the use of Necklace Lie Algebras as an algebraic tool to address the algebraic approach to quantum gravity with huge symmetry [7]. The sticking point has been how to show that this can be related to an emergent spacetime. I will finish this essay by demonstrating a solution to that problem using iterated integration.

**A necklace lie algebra is a lie algebra built from copies of vector spaces strung together in chains**. If the vector space is 2 dimensional you can picture elements of the algebra as necklaces of qubits, and more generally of qudits. These algebras embody the idea of quantised information as a primordial building block. Necklace Lie algebras can take various forms and there is no general definition, but the simplest example is derived from a freely generated associative algebra generated by $d$ independent elements $e_i$. Arbitrary products of these generate new elements which can be written with multiple indices i.e.  $e_i \ldots e_k = e_{i\ldots k}$ A general element of the algebra includes linear sums of these multiplied by components of tensors of any rank, including scalars. Products in the algebra are then just summed tensor products. This constructs an associative algebra which is graded over the non-negative integers where the $r$-graded space is a tensor vector space of dimension $d^r$. The base elements of the algebra can be visualised as open chains where multiplication is concatenation

$$(i - \cdots - k) \times (l - \cdots - n) = l - \cdots k - l - \cdots - n$$

This becomes a necklace Lie algebra $\mathcal{O}_d$ simply by using the commutator as the Lie product.

A related Lie-algebra $\mathcal{C}_d$ can be constructed as the freely generated Lie-algebra from $d$ independent generators. This is graded over positive integers and the dimension of the 1-graded space is again $d$, but since commutators are antisymmetric the dimension of the 2-graded space is $\frac{1}{2}d(d-1)$. In general the dimension of the $r$-graded space (as shown by Ernst Witt who was a student of Emmy Noether) is given by Moreau's necklace-counting function

$$M(d,r) = \sum_{f|r} \mu\left(\frac{r}{f}\right) d^f$$

Where $\mu(x)$ is the number-theoretic Möbius function which is plus or minus one for a square free number depending on whether it has an even or odd number of prime factors. The necklace-counting function is so-called because it counts the number of ways a sequence of numbers from 1, to $d$ can be arranged in a cycle of length $r$ when counting cyclic permutations as equivalent and disallowing cyclic repetitions. This can be observed explicitly in the Lyndon Basis for the free Lie-algebra where each $r$-graded space is described by a basis of Lyndon words which are the lexographically largest representative from the necklace of length $r$. **This tells us that a free Lie algebra $\mathcal{C}_d$ also has the structure of a necklace Lie algebra, but whereas the free associative algebra defines a necklace Lie algebra $\mathcal{O}_d$ over open chains, the free Lie-algebra uses cyclically closed chains**.

The free Lie algebra $\mathcal{C}_d$ is contained in the Lie algebra $\mathcal{O}_d$ but is smaller since the dimensions of its $r$-graded spaces are smaller. However, $\mathcal{C}_d$ can be enlarged to give an associative algebra by constructing its universal enveloping algebra $U(\mathcal{C}_d)$ from products of its elements modulo the Jacobi and anti-symmetry conditions for the Lie-algabra. This is then isomorphic to the free associative algebra. Pictorially, while the free Lie algebra has a basis over single necklaces, its universal enveloping algebra has a basis over unordered collections of necklaces. As a universal enveloping algebra the free associative algebra must also have a commutative graded dual whose product can be shown to be a shuffle product. This is defined as a sum over all ways to merge together two chains of numbers to form new chains such that the original ordering of the two chains is preserved in the merged chain. The dual of the tensor product is a co-product called the de-concatenation product and together with the shuffle product they form a bi-algebra.

**Given this apparent duality between open and closed necklaces in the fundamentals of Lie algebra theory it is tempting to wonder if these can somehow be considered as discrete strings related to string theory**. To make this useful in physics we would need some way of relating these discrete strings to continuous strings.

To understand how this could come about, **an algebra over string states can be constructed**. Take a collection of open strings to be a piecewise continuous mapping $v(s)$ from a real line interval $s \in [0, s_0], 0 \le s_0$ to a $d$-dimensional vector space $V$. Let $S$ be the set of all such string embeddings, then a string state $\varphi \in \Phi$ is a mapping from $S$ to the complex numbers $\mathbb{C}$. $\Phi$ can be made into a commutative algebra using a product defined by the rule $(\varphi_1 \cdot \varphi_2)(v) = \varphi_1(v)\varphi_2(v)$. This algebra can be further extended to a bi-algebra by introducing a co-product $\Delta: \Phi \to \Phi \otimes \Phi$ defined by the rule $(\Delta\varphi)(v_1, v_2) = \varphi(v_1 \parallel v_2)$ where the symbol $\parallel$ represents concatenation of strings.

Now we have two bi-algebras. The first is the free associative algebra which behaves like discrete open chains or closed necklaces. The second is a bi-algebra of continuous open string states. These appear to be very different. At best you might think that the chains could be some kind of discretisation of the continuous strings. Remarkably there is a much more precise relationship between the two. **Using a mapping based on iterated integration we can construct an exact bi-algebra monomorphism $\mathcal{M}$ between them** as follows

$$\left(\mathcal{M}(e_{ij\ldots k})\right)(v) = \int_0^{s_0} ds_1\, v_i(s_1) \int_0^{s_1} ds_2\, v_j(s_2) \ldots \int_0^{s_{r-1}} ds_r\, v_k(s_r)$$

It can now be checked using the rule of partial integration that the shuffle product on the dual of the free associative algebra maps onto ordinary products of the string state functions and that the deconcatenation co-product maps onto the co-product for string states using string concatenation and hence that this mapping is a bi-algebra monomorphism.

In fact this is just one basic example of a more general principle that can be used to map necklace Lie-algebras onto quantum string field states that I hope can be developed to reduce string field theory to purely algebraic terms. In philosophical terms **it can be interpreted as the emergence of string field theories in continuous space and time from algebras based on quantised information**. The full theory is far from complete but I think this example illustrates the potential possibilities.

## *Conclusion*

In his famous essay "It from Bit", Wheeler drew our attention to the fact that we never really measure real numbers. We just answer yes/no questions. Nature's information comes in bits. Other forms of information are human invention. He argued that the thermodynamics of black holes tells us that information is an important basic concept in physics, but is it fundamental or does it emerge from macrophysical phenomena such as statistical physics? **Should we base our theoretical foundation on basic material constructs such as particles and space-time or do these things emerge from the realm of pure information?** Wheeler argued for the latter. But no amount of philosophizing can tell us if this is how the universe works. There is no point in asking where the information comes from, or where it is stored. What we need is a consistent theory built on mathematical logic that accounts for all known observations.

Wheeler had some prophetic words to say about string theory. We must "**Translate the quantum versions of string theory and of Einstein's geometrodynamics from the language of the continuum to the language of bits**" [1]. That was more than a decade before the qubit/black-hole correspondence from string theory showed that the mathematics of quantised information is present at the heart of superstring theory [8].

Wheeler also said that "**Probability like time, is a concept invented by humans**" [1]. This suggests an acatalyptic universe in which nothing is certain, even uncertainty. Von Weizsäcker's multiple quantisation may address this issue.

In quantum mechanics the total probability of the wave function remains constant, normalised to one. When second quantisation is invoked this conservation law **translates to conservation of electric and colour charges and a gauge field is introduced whose flux carries information about the total charges to the boundary in holographic form**. Now a new total probability appears for the quantum field theory. **A further quantisation as envisaged by Weizsäcker would require a new bigger symmetry, a new gauge field and a new flux so that more information is available at the boundary**. This new huge symmetry, required from holography by Noether's theorems, must take the form of higher spins, bosons and fermions. Hence superstring theory or something very like it is required in the bulk volume.

Before even the holographic principle was recognised, I defined necklace Lie algebras as a tentative formulation of string theory nearly twenty years ago. Developments since have only confirmed that the idea makes sense. New work has interpreted the discrete algebras as strings of qubits [9] and related the discrete structures to the continuum through mappings defined by iterated integration. Necklace Lie algebras provide a natural construction to generate a new symmetry from a given one through a process akin to quantisation that can be repeated. I anticipate that multiple quantisation will build on this to realise the version of string theory that Wheeler demanded and Weizsäcker anticipated.